\documentclass[copyright,creativecommons]{eptcs}
\providecommand{\event}{FORECAST 2016} 

\usepackage{xspace}    
\usepackage{amsmath}    
\usepackage{amssymb}    
\usepackage{xspace}    
\usepackage{graphicx}

\newcommand{\csnil}{\mathbf{nil}}
\newcommand{\cskill}{\mathbf{kill}}
\newcommand{\csguard}[2]{[#1]#2}

\newcommand{\csnow}{\mathsf{now}}

\newcommand{\cscpar}{\parallel}
\newcommand{\cszero}{\mathbf{0}}

\newcommand{\broadcastsymbol}{\star}

\newcommand{\csgout}[5]{#2^{#1}[ #3 ]\langle #4\rangle #5 }
\newcommand{\csbout}[4]{\csgout{\broadcastsymbol}{#1}{#2}{#3}{#4}}

\newcommand{\csstore}{\gamma}
\newcommand{\col}{N}
\newcommand{\comp}{C}
\newcommand{\cscomp}[2]{(#1,#2)}

\newcommand{\cssys}[2]{#1~\mathbf{in}~#2}
\newcommand{\csenv}{\mathcal{E}}

\newcommand{\cssel}{\mu}

\newcommand{\guard}{\pi}

\newcommand{\defn}{\stackrel{\scriptscriptstyle\mathrm{def}}{=}}
\newcommand{\stime}{\mathit{stime}}
\newcommand{\average}{\mathit{average}}
\newcommand{\total}{\mathit{total}}
\newcommand{\countr}{\mathit{count}}
\newcommand{\csmy}{\mathsf{my}}

\title{Modelling movement for collective adaptive systems with CARMA}

\author{Natalia Zo\'n \qquad\qquad\quad Vashti Galpin \qquad\qquad\quad
Stephen Gilmore
\institute{Laboratory for Foundations of Computer Science\\
School of Informatics, University of Edinburgh}
\email{s1478292@sms.ed.ac.uk \qquad Vashti.Galpin@ed.ac.uk \qquad
       Stephen.Gilmore@ed.ac.uk}
}
\def\titlerunning{Modelling movement for collective adaptive systems}
\def\authorrunning{N. Zo\'n, V. Galpin and S. Gilmore}

\newcommand{\Carma}{\textsc{Carma}\xspace}
\begin{document}
\maketitle

\begin{abstract}
Space and movement through space play an important role in many 
collective adaptive systems (CAS). CAS consist of multiple components 
interacting to achieve some goal in a system or environment that can 
change over time. When these components operate in space, then their 
behaviour can be affected by where they are located in that space. 
Examples include the possibility of communication between two components 
located at different points, and rates of movement of a component 
that may be affected by location. The \Carma language and its associated 
software tools can be used to model such systems. In particular, a 
graphical editor for \Carma allows for the specification of spatial structure and 
generation of templates that can be used in a \Carma model with space. 
We demonstrate the use of this tool to experiment with a model of 
pedestrian movement over a network of paths. 
\end{abstract}

\section{Introduction}
Collective adaptive systems consist of multiple components or agents
that interact collaboratively on common goals, and compete to achieve individual goals.
They are characterised by the fact that each component does not
have a global view of the whole system but rather has local information
on which to act. These systems are called \emph{collective} because of the
interaction of many components, and \emph{adaptive} because they respond
to changes in the environment in which they operate. They are often
characterised as having emergent behaviour, that is behaviour which
cannot be predicted in advance by considering the individual components
in isolation from each other. Modelling of CAS is crucial because
it is difficult to understand the behaviour of the overall system
just by inspecting the behaviour of the components. Modelling allows
us to experiment with the system before implementation and deployment.
This paper considers the \Carma language which has been developed
specifically for the modelling of CAS \cite{CARMA-tutorial}
with a particular focus on smart
city concerns such as smart transport and smart energy grids.

The use of \emph{local} versus \emph{global} above suggests that
space may play an important role in CAS. While this is not necessarily
true of all CAS since the distinction between local and global may
be logical or virtual rather than physical, it is true for many
CAS, and hence this is an important part of understanding their
behaviour. This paper focusses on a spatial example and illustrates how
\Carma can be used to model this example.

We consider the example of pedestrians moving over a network of paths.
This could be a specific part of a city, a pedestrianised network of
lanes, or paths through a large park. The defining feature of our
example is that there are essentially two groups of pedestrians that
start on opposite sides of the network who wish to traverse the paths to
get to the side opposite to where they started. This scenario could
arise in a city where there are two train stations on opposite sides of
the central business district serving the eastern and the western suburbs of
the city, and a number of people who commute from the west work close to
the east station and vice versa. During rush hour in the morning and
afternoon, people want to traverse the park or lanes as fast as
possible so that they are not late, and we wish to investigate what
features enable these pedestrians to pass through the network
efficiently.  If there are multiple paths, it would seem in advance
that it makes sense to use some paths for one direction and other
paths for the other direction.  This raises the question of what
routing or information such as signs is sufficient for the two
groups of pedestrians to separate out onto different paths.
This paper presents an initial investigation into the modelling of
this scenario, and we demonstrate how this can be achieved using
\Carma, its Eclipse Plug-In and its Graphical Eclipse Plug-in,
considering different possibilities for the network.

The paper starts with a brief discussion of the \Carma language
before describing the modelling of the scenario in more detail and
presentation of our experiments and results from various networks,
followed by conclusions and discussion of future work.

\section{\Carma}

The \Carma process calculus has been developed specifically for the modelling of
collective adaptive systems and a full description of the language can be
found in~\cite{CARMA-tutorial}. Here, we give a brief outline. A \Carma model consists
of a collective $\col$ and the environment $\csenv$ in which it operates, using the
syntax
$\cssys{\col}{\csenv}$. A collective is either a component $\comp$ or
collectives in parallel $\col\cscpar \col$. Each component is either
null, $\cszero$, or a combination of behaviour described by a
process $P$ and a store of attributes $\csstore$, denoted by~$\cscomp{P}{\csstore}$.
We use function notation to denote store access, thus if $\gamma=\{x\mapsto v\}$ then 
$\gamma(x)=v$.

Prefix, constants, choice and parallel composition can be expressed
in the standard manner by defining $P$ appropriately. Additionally,
there is the $\csnil$ process which does nothing, the $\cskill$
process which results in the component being removed from the
collective, and the option of prefixing a process with a predicate
$\csguard{\guard}{P}$, in which case the process $P$ can only proceed
if the predicate $\guard$ evaluates to \emph{true} using the values of the attributes
in the component's store $\csstore$.  To improve readability we sometimes parenthesise 
the process expression $P$, writing this term as $\csguard{\guard}{(P)}$. The
meaning is unchanged.

Process prefixes are rich and permit actions that provide value-passing
unicast and broadcast communication using predicates on the attributes
in the store of the sending and receiving component. Communication
between components will only take place if the predicates evaluate
to \emph{true}.  The value~\emph{false}
indicates that no communication partner is needed.
Furthermore, attribute values can be updated
(probabilistically) on completion of an action. 

Unicast communication
is blocking; the sender cannot output values unless there is a matching
input action which can be performed by another component. In contrast,
broadcast is not blocking, and we can use a specific form with a constant
\emph{false} predicate  (written as~$\perp$ here) to allow components to 
act without interaction with other components, as seen in the example to
follow.

The syntax of a non-blocking broadcast on name~$\alpha$ is 
$\alpha^{\star}[\pi]\langle\vec{v}\rangle\sigma$ where~$\pi$ is a
predicate which must be satisfied by all processes wishing to receive 
this broadcast.   The vector~$\vec{v}$
is a vector of values to be communicated; this vector may be empty.  The 
suffix~$\sigma$ is an \emph{update} of variables in the local store of 
a component.  A component refers to its local store with the prefix~\textsf{my}
(similar to \textsf{this} in Java) so an update to store the value of~$x$ as 
the new value of \textsf{my}.$x$ is written as $\{\mathsf{my}.x \leftarrow x\}$.
As an example, the prefix process term $\csbout{\mathsf{move}_{ij}}{\bot}{}
\{\csmy.x\leftarrow i, \csmy.y\leftarrow j\}.\mathit{Ped}$ broadcasts that it is performing
a $\mathsf{move}_{ij}$ activity, updates its local $x$ and $y$ values, and 
continues as the process $\mathit{Ped}$.

The environment contains both the global store and an evolution rule
which returns a tuple of four functions $(\mu_p, \mu_w, \mu_r, \mu_u)$
known as the \emph{evaluation context}.  Communication between sender~$s$
and receiver~$r$ on activity~$\alpha$ has both an associated \emph{probability} (determined
by $\mu_p$) and a \emph{weight} (determined by~$\mu_w$).  These functions depend on
activity~$\alpha$  and both the attribute values of the sender (in the store~$\gamma_s$) and the 
attribute values of the receiver (in the store~$\gamma_r$).  The activity \emph{rate} however
depends on only the attribute values in the store of the sender ($\gamma_s$); the 
attribute values of the receiver do not affect the rate at which a communication activity is performed.
Thus the first three functions in the evaluation context determine probabilities, weights and rates
that supply quantitative information about the behaviour of actions. The
fourth function~$\mu_u$ performs global updates, either of the attributes in the
global store or of the collective by adding new components.  These updates include the 
usual initialisation of variables, incrementing counters, or accumulating totals.

The operational semantics of \Carma are defined in \textsc{FuTS} style
\cite{DLLM13}
and define for each model a time-inhomogeneous continuous-time Markov
chain (ICTMC). The behaviour of these ICTMCs can be simulated and the \Carma
Eclipse Plug-in provides this functionality.  The software tools for processing \Carma
models are available from \url{http://quanticol.sourceforge.net}.  More information
about the QUANTICOL project which created these tools is available from 
\url{http://www.quanticol.eu}.

\section{Automatic code generation}

The \Carma Graphical Editor allows the user to
specify the structure of movement in a CAS model by laying out
graphical symbols on a plane \cite{CarPooling}.  The editor generates \Carma code
from the graph which the user has defined.  In addition to normal attributes,
\Carma components which are defined in this way have a set of distinguished 
attributes to specify their current location in space.

Each \Carma component in the model may further
have its mobility restricted to a given set of paths through the
graph defined by the user. Paths in this
context are subgraphs of the user-defined graph consisting of a set of uniform
vertices connected by directed, coloured edges. At any given time
in the system's evolution, the location attributes of a component
instance must be equal to the location of one of the nodes belonging
to the subgraph where that component is
restricted. A component can change its location attributes only if there 
exists a path from its current
node to the
new node, and if this path belongs to the subgraph where the
mobility of this component is restricted.  

In \Carma, functions
are used for storing the information about each subgraph's topology.
Component actions query these functions during the execution of
their predicates, and can modify the component's location attributes
accordingly, in the update block.  For each node which can be
accessed by a particular component type, a movement action  must
be included in the component's behaviour. If the node can be accessed
by a component in more than one state, the action must be specified
separately for each state.  In systems with complex mobility
restriction graphs, topology-defining functions, as well as component
behaviour blocks may require a large number of lines of code. This
code is automatically generated by the \Carma Graphical Editor, freeing
the modeller from the task of manually producing this \Carma model code.

\section{Pedestrian model}

The \Carma model is illustrated in Figures~\ref{fig:model} and
\ref{fig:envcoll}. It assumes that there are two types of pedestrians,
$A$ and $B$, and that $P$ and $Q$ are variables of type pedestrian.
The two $\mathit{Arrival}$ components generate pedestrians at two
different locations (on opposite sides of the graph), and the
pedestrians move from their origin side to the opposite side. Once a 
pedestrian has reached its goal, the count for that type of pedestrian 
is incremented and the time
taken for traversal is added to the total time so that the average
traversal time can be calculated for each pedestrian type.

The model is parameterised by a number of functions that capture the
graph information and are generated automatically as described above.
\begin{itemize}
\item $\mathbf{ExistsPath}(P,x,y,i,j)$ is a Boolean function that
determines if an edge exists between a pedestrian's current position and
another node in the graph, hence a $\mathsf{move}_{ij}^\star$ action can only
occur when such an edge exists.
\item $\mathbf{AtGoal}(P,x,y)$ is a Boolean function that checks if the
pedestrian has reached its goal, hence $\mathsf{fin}^\star$ can only
occur once the destination has been reached. After this the pedestrian
does not move any more.
\item $\mathbf{ArrivalRate}(P)$ determines the arrival rate
for each type of pedestrian.
\item $\mathbf{Start}_x(P)$ and $\mathbf{Start}_y(P)$ define the
initial location of a new pedestrian depending on its type.
\end{itemize}
A function that is not directly related to the graph structure is
$\mathbf{MoveRate}(P,x,y,i,j,\ldots)$ which determines the rate of movement
along a particular edge, and can take additional parameters that can
affect this rate such as the current count of other pedestrians of the
same or different type. We use the following definition that uses the
numbers of pedestrians of the other type at the target node to reduce
the movement rate.
\[ \mathbf{MoveRate}(P,x,y,i,j,A_{ij},B_{ij})
=\begin{cases}
 \mathit{move_A}/(B_{ij}+1) & \text{if } P=A \\
 \mathit{move_B}/(A_{ij}+1) & \text{if } P=B \\
\end{cases} \]
where $A_{ij}$ are the number of $A$ pedestrians at the target node and
$B_{ij}$ are the number of $B$ pedestrians at the target node, and
$\mathit{move_Q}$ is a basic movement rate for each pedestrian type.


\begin{figure}[t]

{\small
$\begin{array}{rcl}
\multicolumn{3}{l}{\textbf{Store of $\mathit{Pedestrian}$ component:}} \\[2ex]
P & & \text{pedestrian type --- an enumeration with values $A$ and $B$} \\
x & & \text{current $x$ coordinate} \\
y & & \text{current $y$ coordinate} \\
\stime & & \text{time of arrival} \\[2ex]
\multicolumn{3}{l}{\textbf{Behaviour of $\mathit{Pedestrian}$
component:}} \\[2ex]
\mathit{Ped} & \defn &
  \sum_{(i,j)\in V} \big[ \mathbf{ExistsPath}(P,x,y,i,j) \big]
  \big(\csbout{\mathsf{move}_{ij}}{\bot}{}
  \{\csmy.x\leftarrow i, \csmy.y\leftarrow j\}.Ped\big)\\[1.25ex]
  &  & \hspace{50.125pt} {} +
  \big[ \mathbf{AtGoal}(P,x,y) \big] \big(\csbout{\mathsf{fin}}{\bot}{}.\csnil \big)\\[2ex]
\multicolumn{3}{l}{\textbf{Initial state of $\mathit{Pedestrian}$
component:} \quad\mathit{Ped}} \\
\\
\hline
\\
\multicolumn{3}{l}{\textbf{Store of $\mathit{Arrival}$ component:}} \\[2ex]
P & & \text{pedestrian type} \\[2ex]
\multicolumn{3}{l}{\textbf{Behaviour of $\mathit{Arrival}$ component:}} \\[2ex]
Arr & \defn  & \csbout{\mathsf{arrive}}{\bot}{}.Arr \\[2ex]
\multicolumn{3}{l}{\textbf{Initial state of $\mathit{Arrival}$
component:} \quad\mathit{Arr}} \\
\end{array}$}
\caption{The $\mathit{Pedestrian}$ and $\mathit{Arrival}$ components}
\label{fig:model}
\end{figure}

\begin{figure}[htb]
{\small
$\begin{array}{lcl}
\multicolumn{3}{l}
{\textbf{Constants:}} \\
$V$ & & \text{set of coordinate pairs representing nodes in the graph} \\[2ex]
{\textbf{Measures:}} \\
\average_P & & \text{average time for traversal by pedestrians of type
$P$} \\[2ex]
{\textbf{Global store:}} \\
\countr_P & & \text{number of $P$ pedestrians to complete traversal} \\
\total_P & & \text{total time for all completed $P$ pedestrian traversals} \\[2ex]
\multicolumn{3}{l}{\textbf{Evolution rule functions:}} \\

\cssel_{p}(\gamma_{s},\gamma_{r},\alpha) & = & \:\:\:\:\:\:\: 1 \\
\cssel_{w}(\gamma_{s},\gamma_{r},\alpha) & = & \:\:\:\:\:\:\: 1 \\
\\
\cssel_{r}(\gamma_{s},\alpha) & = &
\begin{cases}
\mathbf{ArrivalRate}\big(\gamma_s(P)\big) & \qquad
		\text{if } \alpha=\mathsf{arrive}^\broadcastsymbol \\
\mathbf{MoveRate}\big(\gamma_s(P),\gamma_s(x),\gamma_s(y),i,j,\ldots\big) & \qquad
		\text{if } \alpha=\mathsf{move}_{ij}^\broadcastsymbol \\
\lambda_\mathit{fast} & \qquad \mbox{otherwise}
\end{cases} \\
\\
\cssel_{u}(\gamma_{s},\alpha) & = &
\begin{cases}
\multicolumn{2}{l}{\!\!\!\!\big\{\big\},\big(\mathit{Pedestrian},
\{P \leftarrow \gamma_s(P),
x \leftarrow \mathbf{Start}_x(\gamma_s(P)),
y \leftarrow \mathbf{Start}_y(\gamma_s(P)),\stime \leftarrow \csnow\}\big)} \\
& 		\text{if } \alpha=\mathsf{arrive}^{\star} \\

\multicolumn{2}{l}{\!\!\!\!\big\{\countr_{\gamma_s(P)} \leftarrow
\countr_{\gamma_s(P)}+1, 
\total_{\gamma_s(P)} \leftarrow \total_{\gamma_s(P)} 
+ (\csnow - \gamma_s(stime)) \big\},0} \\ 
\hspace*{6.0cm}
&		\text{if } \alpha=\mathsf{fin}^{\star} \\

\big\{\big\},0 & \mbox{otherwise}

\end{cases} \\
\end{array}$

$\begin{array}{rcl}\\
\multicolumn{3}{l}{\textbf{Collective:}} \\[2ex]
\mathit{PedAB} & \defn & \big(\mathit{Arrival},
\{P\mapsto A\}\big) \cscpar \big(\mathit{Arrival},
\{P\mapsto B\}\big)\\
\end{array}$}
\caption{Environment and collective}
\label{fig:envcoll}
\end{figure}

Figure~\ref{fig:envcoll} specifies the four functions $(\mu_p, \mu_w, \mu_r, \mu_u)$
known as the evaluation context.  Probabilities and weights on activities are not used 
in this model so the $\mu_p$ and $\mu_w$ functions are trivially constant functions. 

\section{Model instances}
\label{sec:instance}

\begin{figure}[!ht]
  \centering
    \begin{tabular}{cc}
    \includegraphics[width=150pt,height=70pt,trim=20 10 15 10,clip]{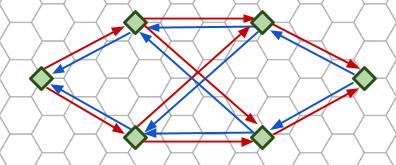}&
    \includegraphics[width=220pt,height=70pt,trim=10 10 12 10,clip]{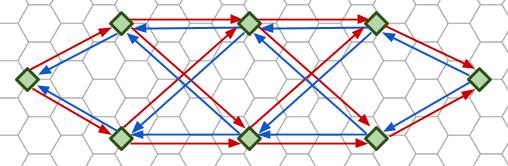}\\
    $1\times1$ & $1\times2$\\
    \\
    \includegraphics[width=150pt,height=120pt,trim=20 20 15 15,clip]{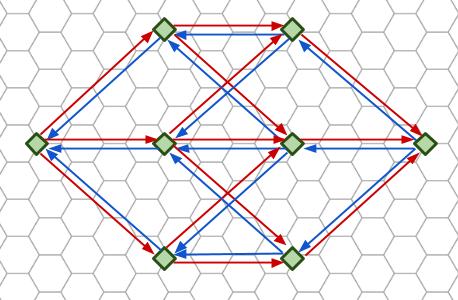}&
    \includegraphics[width=220pt,height=120pt,trim=20 10 10 20,clip]{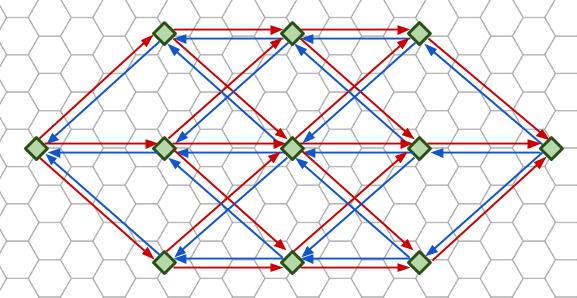}\\
    $2\times1$ & $2\times2$\\
    \end{tabular}
      \caption{\label{fig:modelAB}Four model instances of increasing size and complexity.}
\end{figure}

Four instances of this \Carma model are shown in Figure~\ref{fig:modelAB}.
These show instantiations of the general \Carma model from the 
previous section with increasing size and shape complexity.  The 
central repeating features of the path network are the cross-bars in the 
centre of the network.  In the simplest instance we have only one
cross-bar and we describe this instance as having height~1 and width~1, representing
it as instance~$1 \times 1$.  As the cross-bar structure is repeated we have
instances~$1 \times 2$, $2 \times 1$, and~$2 \times 2$, depending where
the additional structure is added into the network.  

An increase in the width of the network has the obvious consequence that 
journeys across the network take longer.  An increase in the height of the
network has the consequence that pedestrians are offered an increased
choice of routes, with the implicit consequence that individual paths are 
less congested (because there are more of them on offer).  The edges are 
equally long and thus the time to traverse them is the same under 
comparable conditions.

In each instance of the network of paths there are two
sub-networks which restrain the movement of the pedestrians of type~$A$
and type~$B$\@.  Pedestrians of type~$A$ are restricted to the red
sub-network and must cross the network from left to right.  Pedestrians
of type~$B$ are restricted to the blue sub-network and must cross the
network from right to left.  The networks illustrated in Figure~\ref{fig:modelAB} 
are symmetric but this is of no particular significance and it would pose no 
difficulty to work with networks which were not symmetric.

These graphs were drawn in the \Carma
Graphical Editor and \Carma code was generated from it, including
all necessary instances of the \textbf{ExistsPath}, \textbf{AtGoal},
and \textbf{ArrivalRate} functions and applications of these in
predicate guards on processes.

\section{Analysis and results}
\label{sec:analysis.and.results}

We analysed our \Carma model using the \Carma Eclipse
Plugin~\cite{CARMA-tutorial}.  The \Carma Eclipse Plugin provides
a helpful syntax-aware editor for the \emph{\Carma Specification
Language}, implemented in the XText editor framework.  The \Carma
Specification Language provides a wrapper around the \Carma process
calculus adding non-essential (but useful) features such as data
types and data structures, functions, and the ability to specify
real-valued \emph{measures} of interest over the model.  In some
modelling languages measures of interest or Markov reward structures
are defined externally to the model but in \Carma and languages
such as CASPA~\cite{DBLP:conf/forte/KuntzSW04}, PRISM~\cite{KNP11}
and ProPPA~\cite{Georgoulas2014}, the specification of measures of
interest and reward structures is included in the modelling language
itself.

Given a \Carma specification, the \Carma Eclipse Plug-in compiles
the model into a set of Java classes which are linked with the
\Carma  simulator classes to provide a custom simulator for this
model.  The compiled Java code is executed to compute the measures
of interest from an ensemble of simulation runs.  The \Carma simulator
uses a \emph{kinetic Monte-Carlo} algorithm to select the next
simulation event to fire and draws from the appropriate weighted
random number distribution to determine the duration of the event.
The simulation state is updated as specified by the event which was
fired and the simulation proceeds forward until a pre-specified
simulation stop time is reached.

The measure functions defined by the modeller are passed into the
simulation environment and provide a view onto the raw simulation
results at intervals which are specified by the modeller.  The
Apache Commons Math Library is used within the Plug-in to perform
statistical analysis of the data.  The \emph{Simulation Laboratory
View} provided by the \Carma Eclipse Plug-in acts as an electronic
laboratory notebook, recording details of the simulation studies
which have been performed.

The \Carma Eclipse Plugin and the \Carma Graphical Editor are available 
from the SourceForge website at
\url{http://quanticol.sourceforge.net}.  After installation they can
be kept up-to-date using the standard mechanism in Eclipse to check
for updates.

\subsection{Design of experiments}

We designed a suite of experiments to explore the behaviour of the model.  To provide a baseline for average travel time we investigated the travel time in the presence of only one type of pedestrian (thereby giving a model which has no congestion).  Thereafter we investigated the models with congestion in the presence or absence of pedestrian routing.  When routing is present, only one starting route has a non-zero rate, and the non-zero rate is assigned in order to direct pedestrians away from each other.

\begin{figure}[!ht]
\begin{center}
{\begin{minipage}[t]{200pt}
\vspace{0pt}
\centering
 \begin{tabular}{||c c c c||}
 \hline
 Model & Nodes & Connections & LoC \\ [0.5ex]
 \hline\hline
 1x1 & 6 & 8 & 208 \\
 \hline
 1x2 & 8 & 12 & 248 \\
 \hline
 1x3 & 10 & 16 & 288 \\
 \hline
 2x1 & 8 & 13 & 258 \\
 \hline
 2x2 & 11 & 20 & 328 \\
 \hline
 2x3 & 14 & 27 & 398 \\
 \hline
 3x1 & 10 & 18 & 308 \\
 \hline
 3x2 & 14 & 28 & 408 \\
 \hline
 3x3 & 18 & 38 & 508 \\
 \hline
\end{tabular}
\end{minipage}}%
{\begin{minipage}[t]{170pt}
\vspace{0pt}
\centering
  \framebox{\includegraphics[height=138pt]{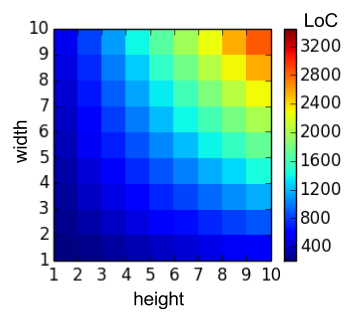}}
\end{minipage}}
\end{center}
      \caption{\label{fig:linesOfCode} The number of lines of \Carma code per model structure.  Left, for small values of width and height.  Right, for larger values of width and height.}
\end{figure}

We used the \Carma Graphical Editor to automatically create \Carma code models. Fig. \ref{fig:linesOfCode} shows how the number of lines of \Carma code grows with model structure complexity.

\subsection{Analysis}

\begin{figure}[htbp]
\begin{center}
\includegraphics[width=0.6\textwidth]{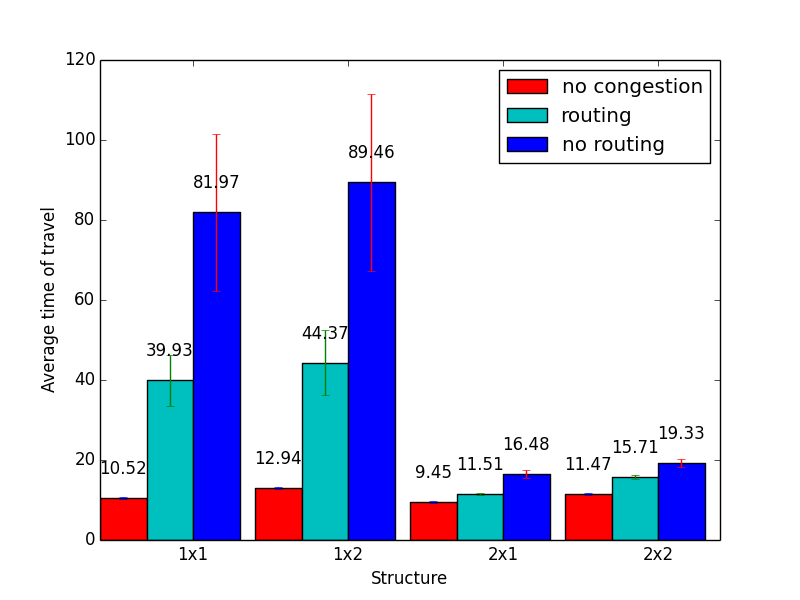}
\caption{\label{fig:results}Average travel time results from the experiments on structure and network usage.}
\end{center}
\end{figure}

The results from our experiments are presented in Figure~\ref{fig:results}.  We have three results (no congestion, routing, and no routing) for each of the four model instances considered ($1\times1$, $1\times2$, $2\times1$, and $2\times2$).
An inspection of the results shows that, unsurprisingly, for any structure the best average travel times are obtained when there is no congestion in the network.  As anticipated, networks with greater height have lower average travel times because they have greater capacity, due to the inclusion of additional routes (thus~$2\times1$ results are better than~$1\times1$ results, and $2\times2$ results are better than~$1\times2$ results).

Finally, we see that routing is always advantageous, especially so in the case of narrow networks where congestion in experienced most (i.e. in the~$1\times1$ structure and the~$1\times2$ structure).

%
%
%


\section{Related work}

The \Carma language provides high-level language constructs for describing communicating processes.  The language has a stochastic semantics expressed in terms of continuous-time Markov chains.  The language contains some features which are familiar from languages such as Bio-PEPA~\cite{DBLP:journals/tcs/CiocchettaH09}, PRISM~\cite{KNP11} and the Attributed~$\pi$-calculus~\cite{John2008}.  In this section we compare \Carma to these established modelling languages and highlight differences in approach between them.  

Each of the languages considered here has the potential to be used to model stochastic systems with mobile populations of individuals but language design decisions, the choice of language features, and underlying analysis mechanisms can make one of the languages better-suited for a particular modelling problem than the others.  
As examples of modelled systems, Bio-PEPA has been used to model scenarios where safe movement of people is an important factor in systems
including \emph{emergency egress}~\cite{DBLP:journals/fac/MassinkLBHH12} 
and \emph{crowd formation and movement}~\cite{DBLP:conf/coordination/BortolussiLM13}.
PRISM has been used to model \emph{dynamic power management controllers}~\cite{NPK+02} 
and \emph{human-in-the-loop UAV mission planning}~\cite{Feng:2015:CSA:2735960.2735973}.
The Attributed~$\pi$-calculus has been used to model \emph{spatial movement in phototaxis}~\cite{John2008}, and \emph{cooperative protein binding in gene regulation}~\cite{John:2010:APP:2172311.2172313}.
\Carma has been used to model a number of spatial CAS including \emph{carpooling}~\cite{CarPooling}, \emph{taxi movement}~\cite{HillHL:15a} and \emph{ambulance
deployment}~\cite{VashtiAmbulance}.  

Inter-process communication in \Carma is \emph{attribute-based}; communication partners are determined dynamically as the model evolves through state-to-state transitions.  Communication in the Attributed~$\pi$-calculus is similarly dynamic.  In contrast, the communication partners of Bio-PEPA and PRISM components are determined statically, and do not change as state-to-state transitions occur.  Additionally, \Carma and the Attributed~$\pi$-calculus support value-passing communication whereas the Bio-PEPA and PRISM languages do not.

The primary analysis method for \Carma models is simulation.  This is also the case for Bio-PEPA and the Attributed~$\pi$-calculus whereas analysis of PRISM models is typically through probabilistic model-checking.

The \Carma language and the PRISM language are explicitly-typed.  Types such as boolean, integer and real are ascribed to variables in the language by the modeller, or inferred by the language type-checker.  In contrast, types in Bio-PEPA and the Attributed~$\pi$ calculus are implicit.  Explicitly-typed languages can make the modeller's intentions more obvious, when, for example, expecting to receive an initial integer value instead of a real value.

\Carma provides guarded process definitions (used in a similar way to the guarded commands found in the PRISM language; the Attributed $\pi$-calculus does not support these directly; Bio-PEPA has no boolean expressions at all).  Guarded process definitions allow declarative descriptions of the relationships between locations in a network and we have used this description mechanism comprehensively here.

In common with PRISM, \Carma provides strong support for encapsulation, with variable declarations being local to an enclosing structure (in PRISM this is a module, whereas in \Carma it is a component).  A structuring mechanism such as this is not found in Bio-PEPA or the Attributed~$\pi$-calculus where declarations of rate functions, channel names, process definitions or species definitions have global scope.

Differently from the other languages considered here, \Carma treats location and space as an aspect of a model which can be described separately from the detailed model dynamics.  Through the provision of a graphical editor for \Carma, space, location, and connectivity can be treated separately from logic, communication, and synchronisation.  This separation of concerns may make it easier to maintain a model of a system when the spatial structure of the system changes.

\section{Conclusions and future work}

We have demonstrated a simple model of pedestrian movement over a
number of different graphs, to illustrate the modelling of spatial aspects of
CAS\@. The \Carma Graphical Editor allowed
us to automatically generate the \Carma code for different
networks which simplified the task, and allowed our pedestrian
components to be generic in nature. Our initial experiments
have considered situations with and without congestion as well as
with and without explicit routing of pedestrians as they enter the
network. 

There are many directions for future work. For example, another
group of pedestrians could also be introduced, namely, tourists,
and the focus would be on efficient traversal of the network during
afternoon rush hours when commuters want to get home quickly and tourists
wish to sightsee, and hence move slowly. We are also interested in
identifying when the model shows emergent behaviour, in the sense that
different groups of pedestrian use different paths through the network
in response to environmental cues such as information about congestion or
routing suggestions (rather than explicit routing).

\paragraph{Acknowledgements:} This work is supported by the EU project
QUANTICOL, 600708.  We thank the anonymous reviewers for many helpful 
comments which encouraged us to improve the paper.

\bibliographystyle{eptcs}
\bibliography{revisedBibliography}

\begin{thebibliography}{10}
\providecommand{\bibitemdeclare}[2]{}
\providecommand{\surnamestart}{}
\providecommand{\surnameend}{}
\providecommand{\urlprefix}{Available at }
\providecommand{\url}[1]{\texttt{#1}}
\providecommand{\href}[2]{\texttt{#2}}
\providecommand{\urlalt}[2]{\href{#1}{#2}}
\providecommand{\doi}[1]{doi:\urlalt{http://dx.doi.org/#1}{#1}}
\providecommand{\bibinfo}[2]{#2}

\bibitemdeclare{inproceedings}{DBLP:conf/coordination/BortolussiLM13}
\bibitem{DBLP:conf/coordination/BortolussiLM13}
\bibinfo{author}{L.~\surnamestart Bortolussi\surnameend},
  \bibinfo{author}{D.~\surnamestart Latella\surnameend} \&
  \bibinfo{author}{M.~\surnamestart Massink\surnameend} (\bibinfo{year}{2013}):
  \emph{\bibinfo{title}{Stochastic Process Algebra and Stability Analysis of
  Collective Systems}}.
\newblock In \bibinfo{editor}{R.~De \surnamestart Nicola\surnameend} \&
  \bibinfo{editor}{C.~\surnamestart Julien\surnameend}, editors: {\sl
  \bibinfo{booktitle}{Coordination Models and Languages, 15th International
  Conference, {COORDINATION} 2013, Held as Part of the 8th International
  Federated Conference on Distributed Computing Techniques, DisCoTec 2013,
  Florence, Italy, June 3-5, 2013. Proceedings}}, {\sl \bibinfo{series}{Lecture
  Notes in Computer Science}} \bibinfo{volume}{7890},
  \bibinfo{publisher}{Springer}, pp. \bibinfo{pages}{1--15},
  \doi{10.1007/978-3-642-38493-6\_1}.

\bibitemdeclare{article}{DBLP:journals/tcs/CiocchettaH09}
\bibitem{DBLP:journals/tcs/CiocchettaH09}
\bibinfo{author}{F.~\surnamestart Ciocchetta\surnameend} \&
  \bibinfo{author}{J.~\surnamestart Hillston\surnameend}
  (\bibinfo{year}{2009}): \emph{\bibinfo{title}{Bio-PEPA: {A} framework for the
  modelling and analysis of biological systems}}.
\newblock {\sl \bibinfo{journal}{Theor. Comput. Sci.}}
  \bibinfo{volume}{410}(\bibinfo{number}{33-34}), pp.
  \bibinfo{pages}{3065--3084}, \doi{10.1016/j.tcs.2009.02.037}.

\bibitemdeclare{article}{DLLM13}
\bibitem{DLLM13}
\bibinfo{author}{R.~\surnamestart {De Nicola}\surnameend},
  \bibinfo{author}{D.~\surnamestart Latella\surnameend},
  \bibinfo{author}{M.~\surnamestart Loreti\surnameend} \&
  \bibinfo{author}{M.~\surnamestart Massink\surnameend} (\bibinfo{year}{2013}):
  \emph{\bibinfo{title}{A uniform definition of stochastic process calculi}}.
\newblock {\sl \bibinfo{journal}{{ACM} Computing Surveys}}
  \bibinfo{volume}{46}, p.~\bibinfo{pages}{5}, \doi{10.1145/2522968.2522973}.

\bibitemdeclare{inproceedings}{Feng:2015:CSA:2735960.2735973}
\bibitem{Feng:2015:CSA:2735960.2735973}
\bibinfo{author}{L.~\surnamestart Feng\surnameend},
  \bibinfo{author}{C.~\surnamestart Wiltsche\surnameend},
  \bibinfo{author}{L.~\surnamestart Humphrey\surnameend} \&
  \bibinfo{author}{U.~\surnamestart Topcu\surnameend} (\bibinfo{year}{2015}):
  \emph{\bibinfo{title}{Controller Synthesis for Autonomous Systems Interacting
  with Human Operators}}.
\newblock In: {\sl \bibinfo{booktitle}{Proceedings of the ACM/IEEE Sixth
  International Conference on Cyber-Physical Systems}}, \bibinfo{series}{ICCPS
  '15}, \bibinfo{publisher}{ACM}, \bibinfo{address}{New York, NY, USA}, pp.
  \bibinfo{pages}{70--79}, \doi{10.1145/2735960.2735973}.

\bibitemdeclare{inproceedings}{VashtiAmbulance}
\bibitem{VashtiAmbulance}
\bibinfo{author}{V.~\surnamestart Galpin\surnameend} (\bibinfo{year}{2016}):
  \emph{\bibinfo{title}{Modelling Ambulance Deployment with {CARMA}}}.
\newblock In \bibinfo{editor}{A.~\surnamestart Lluch~Lafuente\surnameend} \&
  \bibinfo{editor}{J.~\surnamestart Proen{\c{c}}a\surnameend}, editors: {\sl
  \bibinfo{booktitle}{Coordination Models and Languages: 18th IFIP WG 6.1
  International Conference, COORDINATION 2016, Held as Part of the 11th
  International Federated Conference on Distributed Computing Techniques,
  DisCoTec 2016, Heraklion, Crete, Greece, June 6-9, 2016, Proceedings}},
  \bibinfo{publisher}{Springer}, pp. \bibinfo{pages}{121--137},
  \doi{10.1007/978-3-319-39519-7\_8}.

\bibitemdeclare{inproceedings}{Georgoulas2014}
\bibitem{Georgoulas2014}
\bibinfo{author}{A.~\surnamestart Georgoulas\surnameend},
  \bibinfo{author}{J.~\surnamestart Hillston\surnameend},
  \bibinfo{author}{D.~\surnamestart Milios\surnameend} \&
  \bibinfo{author}{G.~\surnamestart Sanguinetti\surnameend}
  (\bibinfo{year}{2014}): \emph{\bibinfo{title}{Probabilistic Programming
  Process Algebra}}.
\newblock In \bibinfo{editor}{G.~\surnamestart Norman\surnameend} \&
  \bibinfo{editor}{William \surnamestart Sanders\surnameend}, editors: {\sl
  \bibinfo{booktitle}{Quantitative Evaluation of Systems: 11th International
  Conference, QEST 2014, Florence, Italy, September 8-10, 2014. Proceedings}},
  \bibinfo{publisher}{Springer}, pp. \bibinfo{pages}{249--264},
  \doi{10.1007/978-3-319-10696-0\_21}.

\bibitemdeclare{inproceedings}{HillHL:15a}
\bibitem{HillHL:15a}
\bibinfo{author}{J.~\surnamestart Hillston\surnameend} \&
  \bibinfo{author}{M.~\surnamestart Loreti\surnameend} (\bibinfo{year}{2015}):
  \emph{\bibinfo{title}{Specification and Analysis of Open-Ended Systems with
  {\Carma}}}.
\newblock In: {\sl \bibinfo{booktitle}{Fourth International Workshop on Agent
  Environments for Multi-Agent Systems, Revised Selected and Invited Papers
  ({E4MAS} 2014)}}, \bibinfo{series}{LNCS 9068}, \bibinfo{publisher}{Springer},
  pp. \bibinfo{pages}{95--116}, \doi{10.1007/978-3-319-23850-0\_7}.

\bibitemdeclare{inproceedings}{John2008}
\bibitem{John2008}
\bibinfo{author}{M.~\surnamestart John\surnameend},
  \bibinfo{author}{C.~\surnamestart Lhoussaine\surnameend},
  \bibinfo{author}{J.~\surnamestart Niehren\surnameend} \&
  \bibinfo{author}{A.M. \surnamestart Uhrmacher\surnameend}
  (\bibinfo{year}{2008}): \emph{\bibinfo{title}{The Attributed Pi Calculus}}.
\newblock In \bibinfo{editor}{M.~\surnamestart Heiner\surnameend} \&
  \bibinfo{editor}{A.M. \surnamestart Uhrmacher\surnameend}, editors: {\sl
  \bibinfo{booktitle}{Computational Methods in Systems Biology: 6th
  International Conference CMSB 2008, Rostock, Germany, October 12-15, 2008.
  Proceedings}}, \bibinfo{publisher}{Springer Berlin Heidelberg},
  \bibinfo{address}{Berlin, Heidelberg}, pp. \bibinfo{pages}{83--102},
  \doi{10.1007/978-3-540-88562-7\_10}.

\bibitemdeclare{inproceedings}{John:2010:APP:2172311.2172313}
\bibitem{John:2010:APP:2172311.2172313}
\bibinfo{author}{M.~\surnamestart John\surnameend},
  \bibinfo{author}{C.~\surnamestart Lhoussaine\surnameend},
  \bibinfo{author}{J.~\surnamestart Niehren\surnameend} \&
  \bibinfo{author}{A.M. \surnamestart Uhrmacher\surnameend}
  (\bibinfo{year}{2010}): \emph{\bibinfo{title}{The Attributed Pi-calculus with
  Priorities}}.
\newblock In \bibinfo{editor}{C.~\surnamestart Priami\surnameend},
  \bibinfo{editor}{R.~\surnamestart Breitling\surnameend},
  \bibinfo{editor}{D.~\surnamestart Gilbert\surnameend},
  \bibinfo{editor}{M.~\surnamestart Heiner\surnameend} \& \bibinfo{editor}{A.M.
  \surnamestart Uhrmacher\surnameend}, editors: {\sl
  \bibinfo{booktitle}{Transactions on Computational Systems Biology XII}},
  \bibinfo{publisher}{Springer-Verlag}, \bibinfo{address}{Berlin, Heidelberg},
  pp. \bibinfo{pages}{13--76}, \doi{10.1007/978-3-642-11712-1\_2}.

\bibitemdeclare{inproceedings}{DBLP:conf/forte/KuntzSW04}
\bibitem{DBLP:conf/forte/KuntzSW04}
\bibinfo{author}{M.~\surnamestart Kuntz\surnameend},
  \bibinfo{author}{M.~\surnamestart Siegle\surnameend} \&
  \bibinfo{author}{E.~\surnamestart Werner\surnameend} (\bibinfo{year}{2004}):
  \emph{\bibinfo{title}{Symbolic Performance and Dependability Evaluation with
  the Tool {CASPA}}}.
\newblock In \bibinfo{editor}{M.~\surnamestart N{\'{u}}{\~{n}}ez\surnameend},
  \bibinfo{editor}{Z.~\surnamestart Maamar\surnameend}, \bibinfo{editor}{F.L.
  \surnamestart Pelayo\surnameend}, \bibinfo{editor}{K.~\surnamestart
  Pousttchi\surnameend} \& \bibinfo{editor}{F.~\surnamestart Rubio\surnameend},
  editors: {\sl \bibinfo{booktitle}{Applying Formal Methods: Testing,
  Performance and M/ECommerce, {FORTE} 2004 Workshops The FormEMC, EPEW, ITM,
  Toledo, Spain, October 1-2, 2004}}, {\sl \bibinfo{series}{LNCS}}
  \bibinfo{volume}{3236}, \bibinfo{publisher}{Springer}, pp.
  \bibinfo{pages}{293--307}, \doi{10.1007/978-3-540-30233-9\_22}.

\bibitemdeclare{inproceedings}{KNP11}
\bibitem{KNP11}
\bibinfo{author}{M.~\surnamestart Kwiatkowska\surnameend},
  \bibinfo{author}{G.~\surnamestart Norman\surnameend} \&
  \bibinfo{author}{D.~\surnamestart Parker\surnameend} (\bibinfo{year}{2011}):
  \emph{\bibinfo{title}{{PRISM} 4.0: Verification of Probabilistic Real-time
  Systems}}.
\newblock In \bibinfo{editor}{G.~\surnamestart Gopalakrishnan\surnameend} \&
  \bibinfo{editor}{S.~\surnamestart Qadeer\surnameend}, editors: {\sl
  \bibinfo{booktitle}{Proc. 23rd International Conference on Computer Aided
  Verification (CAV'11)}}, {\sl \bibinfo{series}{LNCS}} \bibinfo{volume}{6806},
  \bibinfo{publisher}{Springer}, pp. \bibinfo{pages}{585--591},
  \doi{10.1007/978-3-642-22110-1\_47}.

\bibitemdeclare{inproceedings}{CARMA-tutorial}
\bibitem{CARMA-tutorial}
\bibinfo{author}{M.~\surnamestart Loreti\surnameend} \&
  \bibinfo{author}{J.~\surnamestart Hillston\surnameend}
  (\bibinfo{year}{2016}): \emph{\bibinfo{title}{Modelling and Analysis of
  Collective Adaptive Systems with CARMA and its Tools}}.
\newblock In \bibinfo{editor}{M.~\surnamestart Bernardo\surnameend},
  \bibinfo{editor}{R.~\surnamestart De~Nicola\surnameend} \&
  \bibinfo{editor}{J.~\surnamestart Hillston\surnameend}, editors: {\sl
  \bibinfo{booktitle}{Formal Methods for the Quantitative Evaluation of
  Collective Adaptive Systems: 16th International School on Formal Methods for
  the Design of Computer, Communication, and Software Systems, SFM 2016,
  Bertinoro, Italy, June 20-24, 2016, Advanced Lectures}},
  \bibinfo{publisher}{Springer International Publishing},
  \bibinfo{address}{Cham}, pp. \bibinfo{pages}{83--119},
  \doi{10.1007/978-3-319-34096-8\_4}.

\bibitemdeclare{article}{DBLP:journals/fac/MassinkLBHH12}
\bibitem{DBLP:journals/fac/MassinkLBHH12}
\bibinfo{author}{M.~\surnamestart Massink\surnameend},
  \bibinfo{author}{D.~\surnamestart Latella\surnameend},
  \bibinfo{author}{A.~\surnamestart Bracciali\surnameend},
  \bibinfo{author}{M.~D. \surnamestart Harrison\surnameend} \&
  \bibinfo{author}{J.~\surnamestart Hillston\surnameend}
  (\bibinfo{year}{2012}): \emph{\bibinfo{title}{Scalable context-dependent
  analysis of emergency egress models}}.
\newblock {\sl \bibinfo{journal}{Formal Asp. Comput.}}
  \bibinfo{volume}{24}(\bibinfo{number}{2}), pp. \bibinfo{pages}{267--302},
  \doi{10.1007/s00165-011-0188-1}.

\bibitemdeclare{inproceedings}{NPK+02}
\bibitem{NPK+02}
\bibinfo{author}{G.~\surnamestart Norman\surnameend},
  \bibinfo{author}{D.~\surnamestart Parker\surnameend},
  \bibinfo{author}{M.~\surnamestart Kwiatkowska\surnameend},
  \bibinfo{author}{S.~\surnamestart Shukla\surnameend} \&
  \bibinfo{author}{R.~\surnamestart Gupta\surnameend} (\bibinfo{year}{2002}):
  \emph{\bibinfo{title}{Formal Analysis and Validation of Continuous Time
  {Markov} Chain Based System Level Power Management Strategies}}.
\newblock In \bibinfo{editor}{W.~\surnamestart Rosenstiel\surnameend}, editor:
  {\sl \bibinfo{booktitle}{Proc. 7th Annual IEEE International Workshop on High
  Level Design Validation and Test (HLDVT'02)}}, \bibinfo{publisher}{IEEE
  Computer Society Press}, pp. \bibinfo{pages}{45--50}.

\bibitemdeclare{article}{CarPooling}
\bibitem{CarPooling}
\bibinfo{author}{N.~\surnamestart {Zo\'n}\surnameend},
  \bibinfo{author}{S.~\surnamestart Gilmore\surnameend} \&
  \bibinfo{author}{J.~\surnamestart Hillston\surnameend}
  (\bibinfo{year}{2016}): \emph{\bibinfo{title}{Rigorous graphical modelling of
  movement in Collective Adaptive Systems}}.
\newblock {\sl \bibinfo{journal}{In Proceedings of ISoLA 2016}}.
\newblock \bibinfo{note}{To appear}.

\end{thebibliography}

\end{document}